# Technical Case Study of Privacy-Enhancing Technologies (PETs) for Public Health

Avinash Laddha, Danil Mikhailov, Uyi Stewart

## Purpose

Motivated by its mission to accelerate the social impact sector's ability to optimally leverage data and technology, data.org partnered with Mastercard, Harvard University, OpenDP, Javeriana University, and Sloan Foundation, to run a unique Privacy-Enhancing Technologies (PETs) for Public Health Challenge to unlock innovation in the use of financial transaction data, combined with cross-domain datasets, to develop decision support tools and solutions.

This technical case study presents a detailed analysis of how the combination of the PETs technology, partnership ecosystem, derivative datasets, and analytical and evaluative processes were successfully applied to execute the novel PETs for Public Health Challenge. It also covers several of the challenges faced, solutions implemented, and the project outcomes.

It is envisioned that this technical case study will help professionals and businesses across the private sector, research and academic institutions, and social impact organizations, etc. to understand best practices, learn from our experiences, and where applicable apply similar strategies to their own situations to advance data sharing for social impact.

## Acknowledgment

This technical case study report would not be possible without the contributions of the experts and ecosystem actors who shared their expertise at various stages of the project. We are grateful for the time and effort they have given to this project. To the Project Team and data.org colleagues, especially Avinash Laddha, who pushed this work forward, interpreted insights, and contributed to this final report: thank you!

- Mastercard: Shanna Crumely, Smita Jain, John Derrico, Chapin Flynn
- Data.org: Uyi Stewart Ph.D., Danil Mikhailov Ph.D., Avinash Laddha, Paul Korir Ph.D., Hugo Gruson Ph.D., David Mascarina, Stephanie Earl.
- Harvard University/OpenDP: Salil Vadhan Ph.D., Sharon Ayalde, Yanis Vandecasteele, Vikrant Singhal Ph.D.,
- University of Javeriana: Zulma Cucunuba Ph.D., Nicolas Dominguez, Felipe Abril-Bermúdez, Diana Fajardo,
- Judges: Bubacarr Bah Ph.D., Sean Cavany Ph.D., John Derrico, Jack Fitzsimons, Chapin Flynn, Christine Task Ph.D., Charlie Whittaker Ph.D., Wanrong Zhen Ph.D.
- Global Challenge Awardees: Kris Parag Ph.D., (Imperial College London, UK); Ben Lambert Ph.D., (University of Oxford, UK); Anil Vullikanti Ph.D., Zihan Guan, and Dung Nguyen (University of Virginia); B. Aditya Prakash, and Leo Zhao (Georgia Tech); Ravi Tandon, Payel Bhattacharjee, and Fengwei Tian (University of Arizona); Dmitrii Usynin (Technical University of Munich, Germany); Shubham Kumar, Milan Anand Raj, and Divya Gupta (Indian Institute of Technology Kanpur, India)

Finally, thank you also, to the Sloan Foundation for their financial support to run the challenge and implement this project.

Uyi Stewart, Ph.D.,
Chief Data & Technology Officer, data.org

# Foreword

Back in the Fall of 2020, as I was doing my final interviews for the role of Executive Director of an exciting new nonprofit, founded by the Mastercard Center for Inclusive Growth and Rockefeller Foundation, called data.org, Mike Froman, the then Vice Chairman and President of Strategic Growth at Mastercard, posed me an unexpected challenge: if I got this role, how would I go beyond only looking to Mastercard for funding, but also partner with them to leverage data and their data science talent for social impact? Thinking on my feet, I ended up pitching the project that ended up being the Privacy-Enhancing Technologies (PETs) for Public Health Challenge. The reality, of course, is much more complex than such a simple origin story might suggest. In my previous role as Head of Data & Innovation at the Wellcome Trust, a global funder of health programs, I was already thinking about how new data sources outside of health data could help modellers and public health officials better understand the link between human behaviour, like purchasing and mobility, and the spread of infectious disease. My good friend and colleague, Dr Uyi Stewart was doing the same at the same time at the Gates Foundation, and so were many others, including Mastercard itself.

All of us understood that technology companies like Mastercard were custodians of data that would be immensely useful for research and policy, because of their global reach, scale and local penetration, with the products and solutions they provide being used daily by billions of people. If only these organizations could be persuaded that there were technologies and processes now in place to use such data safely, respecting both the privacy of their customers and the commercial sensitivity of the corporations themselves. Thus, the PETs for Public Health Challenge was born.

Danil Mikhailov, Ph.D., Executive Director, data.org

At Mastercard, we believe when it comes to the data of our customers, partners or individual cardholders: You own it. You control it. You should benefit from the use of it. We protect it. Those statements are at the core of our business, so when we started thinking about how to use data to make a positive impact on communities and individuals around the world, we knew we had to do it the right way. We had to be sure we were adhering to all those principles while finding ways to leverage data for social impact and inspiring other private companies to do the same. So, the question arose… how could we confidently achieve all those goals? Turns out the answer to that question was to connect some of the smartest people in the world, who focus on privacy, encryption, data governance, and data science and epidemiological modelling, with some of the latest privacy enhancing technologies, like differential privacy. The result was the amazing piece of work that you have in front of you. We sincerely hope this work inspires other private sector companies to consider the ways in which data can be used, in a safe, responsible, and transparent manner, to drive meaningful, lasting social impact.

Chapin Flynn, Senior Vice President, Transit and Urban Mobility, Mastercard

# 1. Executive Summary:

Mastercard, a global technology company, sought to explore how data may be used in a privacy preserving environment to support, *at scale,* researchers who are working to advance social impact causes, e.g., public health pandemic management. By applying Differential Privacy (DP) to synthetic transaction data, data.org and collaborators (Harvard University, OpenDP, and Javeriana University) successfully ran a PETs for Public Health Challenge that created six reusable tools/frameworks (see descriptions in section 4) to support data-driven insights for pandemic management. Through this process, we developed an approach on how to generate realistic synthetic transaction data.

Our work has shown that synthetic transactional data combined with mobility and other public health datasets hold significant spatial-temporal information, which can reveal real-time behavioural patterns of populations. When analysed through a differential privacy lens, we found that such data have useful predictive power - actionable insights- for so-called nowcasting (immediate hotspot detection, mobility patterns), and forecasting (identifying future infection rates and contact matrixes) summarized in the following three areas:

1. Diagnostic support for nowcasting, i.e., the ability to explain the present or very near future. Two scenarios to note:
    1.1. Hotspot Detection: Identifying areas of high physical interaction to prioritize resource allocation.
    1.2. Pandemic Adherence Monitoring: Tracking shifts in spending behaviour in response to lockdowns and other interventions.
2. Predictive capacity for forecasting, i.e., predicting future events, trends, or outcomes. Two scenarios to note:
    2.1. Mobility Analysis: Understanding movement patterns to guide quarantine policies and predict infection trends.
    2.2. Contact Matrix Estimation: Estimating interaction frequencies across age groups to assess transmission risks.
3. Development of learning models e.g., epidemic management.
    3.1. Two ML/AI frameworks were developed, designed to incorporate synthetic financial transaction data and various other datasets. This supports the scenarios outlined in points 1 and 2, as well as addressing relevant new use cases.

# 2. Introduction:

Over the past decade, fuelled by advances in Machine Learning (ML), Artificial Intelligence (AI), and other emerging technologies, the use of *private sector data* such as mobile phone data (integrated with other cross-domain datasets) to develop new business and social impact solutions has been steadily growing. For example, it has been used to optimize efficiencies in the gig economy with Uber and Lyft, develop a bustling FinTech market in Nigeria, and to manage endemic diseases linked to the seasonal patterns of dengue in Pakistan and rubella in Kenya.[1]

However, there is now a heightened awareness of the risk associated with poor data management resulting in an escalation in the strictness of laws and regulations that stipulate how data may be handled and combined. Consequently, the advent of data protection regulation in over 71% of countries around the world, with a further 9% with draft legislations underway, has raised the urgency for technologies that can enable data integration and analytics while preserving privacy. Privacy-Enhancing Technologies (PETs) have emerged as a promising solution to navigate this dilemma, enabling secure data integration and analytics for business and social impact applications while preserving individual privacy. Newer PETs methods, such as Differential Privacy (DP), have been shown to provide statistically verifiable protection against identifiability.

While PETs have been around for a while, they have largely been a frontier technology that only a select number of regulators and a few companies in the private sector have had an interest in exploring, not to mention the luxury of being able to do so.

## 3. Problem Statement

Pandemics are an unavoidable global threat that necessitate rapid and informed public health responses. Effective management requires granular, diverse, and timely data to inform public health strategies such as tracking disease spread, mapping mobility patterns, and managing human behavioural changes. However, getting such granular, diverse, and timely data remains a major hurdle. For example, research has shown that traditional data collection methods have limitations and potential pitfalls due to the challenges of combining disparate data sources [2], [3], [4], [5].

To this end, the aims of this project were twofold:

1. Assess the feasibility of applying PETs (such as DP) to synthetic transaction data with a view of informing optimal epidemiological decision-making while preserving privacy in the underlying data, and
2. Collaboratively develop generalizable methods for integrating cross-domain data sets to support practical applications of the PETs framework applied to other use cases.

To achieve these goals, we had to address the following technical challenges:

**3.1. Data Management & Privacy**: How to develop realistic synthetic datasets with a baseline privacy-preserving template

**3.2. Translational Use Cases**: How to generate generalizable use cases for pandemic management

**3.3 Application**: How to apply the use cases to develop applications or solutions from the interpolated & privacy-preserved datasets.

Here's a closer look at each challenge and how we approached them:

**3.1. How to develop a synthetic dataset with a baseline privacy-preserving template?**

Given the inherent sensitivity of financial data, the creation of privacy-preserved synthetic datasets becomes a crucial alternative. We developed a synthetic financial dataset in collaboration with Harvard, incorporating a baseline privacy-preserving template.

The synthetic dataset creation process addressed the sensitive nature of the underlying data by creating a privacy-preserving template including:

- **Mimicking Real Data Format:** The synthetic data replicated the structure of actual transactions without requiring the disclosure of any financial records. Rather, the synthetic dataset was generated using a structured data dictionary designed to mimic realistic weekly merchant transaction behaviour across multiple cities and industries. Key variables included:
    - **ID & Merchant ID**: Unique identifiers ranging from 1 to 10,000.
    - **Date**: Captures end-of-week timestamps from January 1, 2019, to December 27, 2022, with weekly intervals.
    - **Merchant Category**: Represents industry types such as Airlines, Restaurants, Drug Stores/Pharmacies, etc., covering a broad range of consumer sectors.
    - **Merchant Postal Code:** Encodes merchant location, allowing city-level identification using postal code patterns — Medellin ('05'), Bogota ('11'), Brasilia ('-000'), and Santiago (default).
    - **Transaction Type**: Indicates whether the transaction occurred ONLINE or OFFLINE, reflecting the sales channel.
    - **Spend Amount (spendamt)**: The total weekly transaction value per merchant.
    - **Number of Transactions (nb transactions):** The count of individual transactions per merchant per week.

- **Integrating Public Real-World Information:** Postal codes aligned with specific city subdivisions (Medellin, Bogota, Brasilia, Santiago), and merchant categories were inspired by Mastercard's publicly available guidelines [6]. Crucially, real-world COVID-19 data from "Our World in Data" informed the modelling of financial responses to pandemic trends, allowing for realistic simulation without exposing private health or financial details.

- **Leveraging Bayesian Prior Knowledge (general patterns):** We leveraged domain expertise to enhance the realism of the data by calibrating merchant category frequencies (e.g., by applying the mechanism to use #death as a factor in average spend amount, we also tried to simulate realistic changes in earnings or spending behavior based on how many new deaths are happening). We also leveraged Mastercard's reference guide to calibrate Merchant category frequencies to reflect typical industries' distribution within the cities. In addition, we applied a COVID-19 effect multiplier to simulate the pandemic's impact on spending – essentially this means we were able to manage some expected/random penalties (Expected drop in count/sum of amount spent) applied to each merchant category.

- **Applying a Baseline Privacy Algorithm:** Recognizing the potential for even synthetic data to inadvertently reveal patterns related to the real data it was based on, a privacy-preserving algorithm, from OpenDP library, was implemented. This crucial step ensured that the shared dataset offered a measurable level of privacy protection, further mitigating any risk of inference about the original information.

The resulting privacy-preserved synthetic dataset allowed teams to develop and test their epidemic analysis solutions in a safe environment, directly addressing the fact that the underlying financial data would otherwise be completely out of reach for such a challenge. This approach successfully facilitated innovation in data analysis for public health while firmly respecting and upholding data privacy principles.

**3.2. How to generate generalizable use cases for pandemic management?**

Once we created the baseline structure for the dataset, the biggest challenge we encountered was how to generate real-world pandemic use cases to facilitate the use of the dataset for the development of applicable decision support tools. We addressed this challenge in two steps: first, creation of generalizable use cases and second, adapting the synthetic dataset to fit the use cases. These are described below:

Step 1: Creation of Generalizable Use cases:

We created an evaluation framework that was structured around five distinct policy use cases or scenarios based on generalizable data science underpinnings (nowcasting, forecasting, learning models, etc.). Each scenario represents a critical challenge in pandemic management where the integration of synthetic transactional data holds significant promise. These use cases or scenarios defined the specific technical objectives and expected outcomes for tools, solutions, and frameworks that were developed.

**3.2.1 Use Case 1**: Enhancement of Epidemiological Techniques through Synthetic Data Integration

- Technical Challenge: How can synthetic transactional data be seamlessly integrated with established epidemiological techniques to provide a more granular and timely understanding of disease transmission dynamics?
- Expected Outcome: Development of methodologies and tools that demonstrate the effective incorporation of synthetic financial data, protected by differential privacy, into standard epidemiological analyses, leading to improved public health response in real time. This includes the design of agile tools for joint analysis of financial and open datasets.

**3.2.2 Use Case 2**: Inferring Contact Patterns and Constructing the Who Acquires Infection From Whom (WAIFW) Matrix

- Technical Challenge: How can synthetic transactional data be utilized to derive detailed insights into contact patterns across diverse population segments (e.g., age groups, professions, commercial activity involvement)?
- Expected Outcome: Development of techniques that leverage transactional data to inform the who acquires infection from whom (WAIFW) matrix, providing a more accurate representation of infection transmission pathways and enabling more targeted public health interventions.

**3.2.3 Use Case 3**: Real-time Estimation of the Effective Reproduction Number (Rt)

- Technical Challenge: How can synthetic transactional data, specifically in the context of informing contact patterns, contribute to more accurate and timely real-time Rt estimations?
- Expected Outcome: Development of models and algorithms that integrate transactional data-derived contact information to enhance the precision and responsiveness of Rt calculations, a key metric for assessing the trajectory of an epidemic.

**3.2.4 Use Case 4:** Mitigation of Bias in Nowcasting Estimations due to Behavioural Factors

- Technical Challenge: How can synthetic transactional data be employed to identify and correct inherent biases in nowcasting estimations that arise from population behavioural changes and reporting lags?
- Expected Outcome: Development of statistical methodologies that leverage synthetic transactional data as an indicator of behavioural shifts, enabling more accurate real-time predictions and improving overall situational awareness for public health officials.

**3.2.5 Use Case 5:** Integration of Synthetic Transactional Data as a Predictive Feature in Epidemic Forecasting

- Technical Challenge: How can synthetic transactional data be effectively incorporated as a predictive variable in models designed for forecasting epidemic curves (e.g., cases, deaths, Rt), with the goal of enhancing predictive accuracy?
- Expected Outcome: Development and evaluation of forecasting models that demonstrate the added value of synthetic transactional data in improving the reliability and robustness of short-term and potentially long-term epidemic projections.

Step 2: Adapting the Synthetic dataset to the Generalizable Use cases:

We designed specific features in the synthetic dataset to replicate some expected features of the financial time series. We developed a Python code to generate a mock dataset that simulated the financial time series for each merchant category based on ad-hoc parameters correlating them with epidemiological time series extracted from openly available datasets. To ensure consistency with the epidemiological available data and expected trends from each merchant category, we implemented the correlation analysis to refine the data generation process, ensuring better alignment with observed phenomena during the COVID-19 pandemic, particularly regarding correlations with epidemiological metrics.

**3.2.6 Recommendations for Epidemiological Metrics**

- Switching from New Cases to New Deaths:

    Initially, the mock dataset was generated using the incidence of cases as the epidemiological metric to correlate financial transactions with pandemic trends. **We recommended using the incidence of deaths instead of the incidence of cases**. The reason for this is that new cases can be highly influenced by external factors, such as testing availability and transmissibility of variants. For instance, during the Omicron

wave, the high transmissibility resulted in significantly more cases but not a proportional increase in deaths. Death counts are often more reflective of the pandemic's impact on policy decisions and public behaviour, which influences transactional trends. Using deaths helped mitigate unintended drastic shifts in transactional patterns during periods with inflated case numbers.

### 3.2.7 Adjustment of Transaction Sampling Proportions

- City-Specific Proportions:

The number of transactions in each city should reflect the proportion of its population relative to the total population in the dataset. The following population percentages were recommended for scaling:

```
self.populations = {"Medellin": 2569000,
            "Bogota DC": 7181000,
            "Brasilia": 4935000,
            "Santiago": 5561000}
total_population = sum(self.populations.values())
percentages = {key: value/total_population*100 for key, value in self.populations.items()}

{'Medellin': 12.688926207645954,
 'Bogota DC': 35.46873456485232,
 'Brasilia': 24.375185221772202,
 'Santiago': 27.467154005729526}
```

This adjustment ensured aligning the synthetic dataset with realistic demographic distributions.

### 3.2.8 Merchant Category Spending Trends

- Negative observations in the number of transactions:

An initial review of the mock dataset showed the presence of some negative numbers of transactions, which were replaced by 1 to run the correlation analysis; meanwhile, OpenDP's representative resolved the issue by adjusting the number of transactions.

```
> df_mock %>% group_by(nb_transactions) %>% count()
# A tibble: 378 × 2
# Groups:   nb_transactions [378]
   nb_transactions     n
             <int> <int>
 1              -5     1
 2              -3     1
 3              -2     4
 4              -1     5
 5               0    62
 6               1   112
 7               2   260
 8               3   498
 9               4  1043
10               5  1989
# i 368 more rows
# i Use `print(n = ...)` to see more rows
```

Figure 3.1: Frequency Distribution of Spending Amounts at Pandemic

### 3.2.9 Average Spending Amount:

An analysis of the average spending per merchant category at different pandemic stages indicated minimal changes.

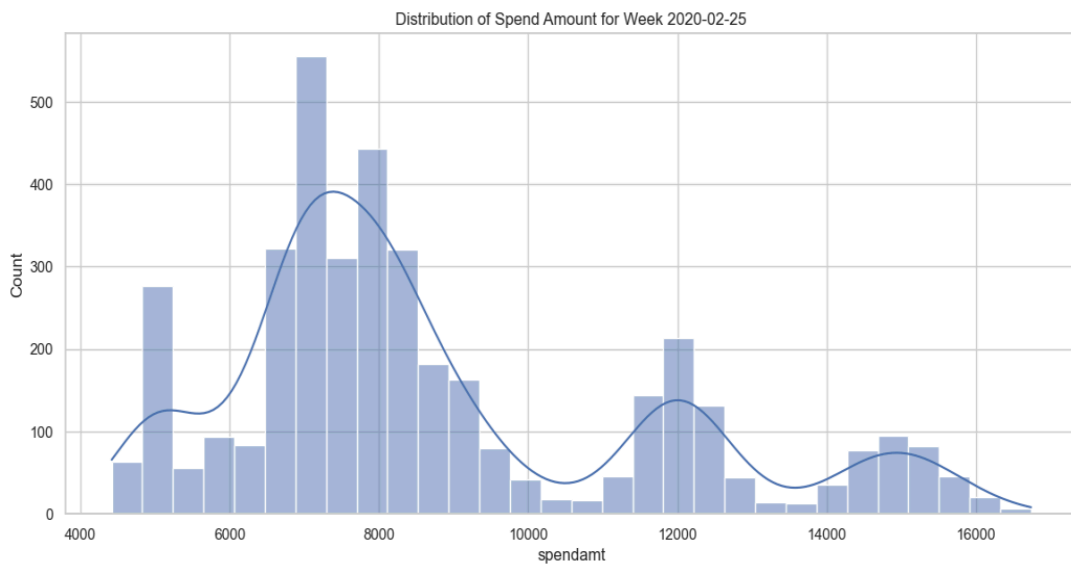

Figure 3.2: Frequency Distribution of Spending Amounts at Pandemic Onset (Illustrative Histogram)

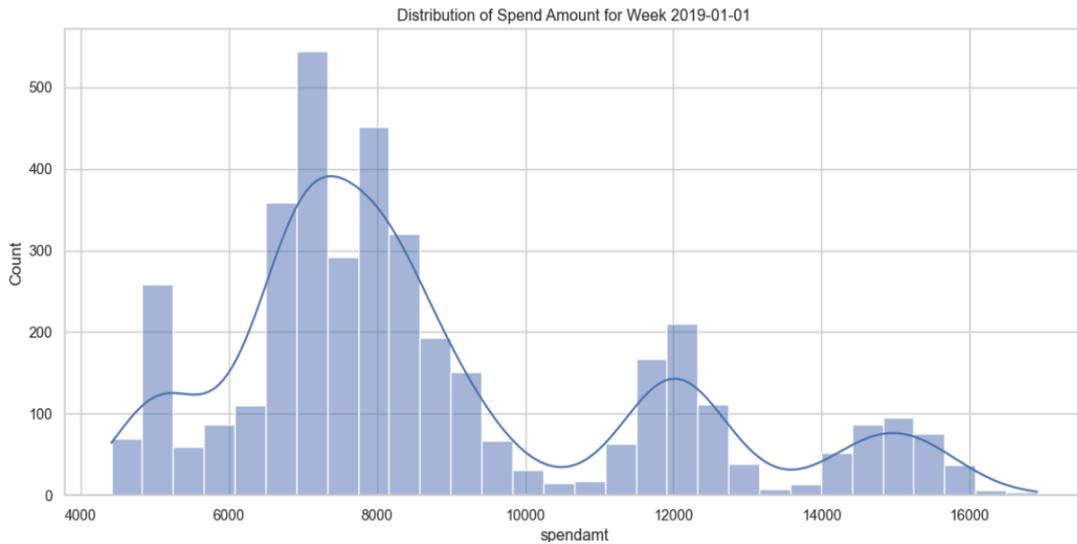

Figure 3.3: Frequency Distribution of Spending Amounts for week 2019-01-01

This stability seemed to have been due to the choice of multipliers for each category and the static nature of typical spending amounts. We suggested reviewing and adjusting these multipliers to better reflect dynamic changes during the pandemic.

### 3.2.10 Frequency Distribution:

A similar issue was observed in the frequency distribution of spending amounts at the pandemic's onset, as exemplified by the following histograms, which correspond to different weeks (2020-02-25 and 2022-12-27 respectively).

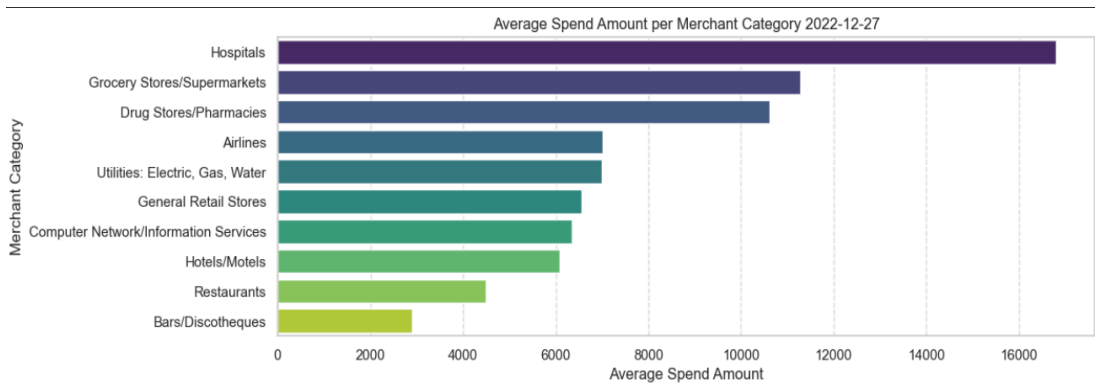

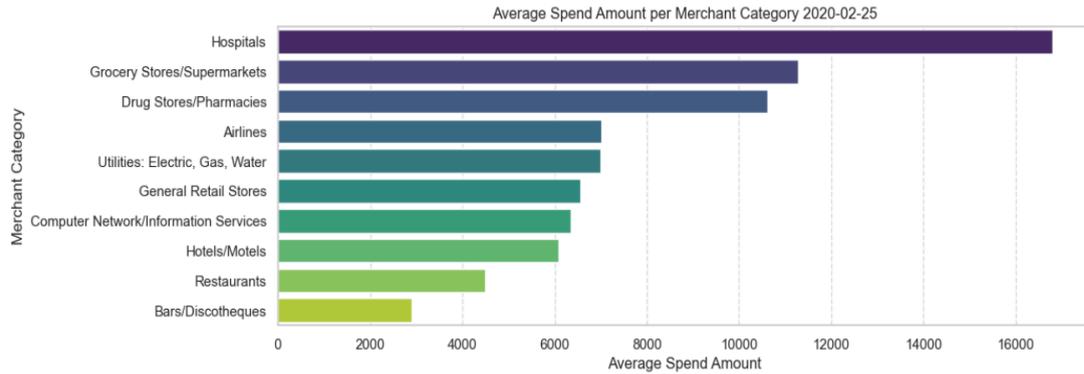

Figure 3.4 and 3.5: Frequency Distribution of Spending Amounts for Week 2020-02-25 (Pandemic Onset) and Week 2022-12-27 (Later Stage)

While clustering effects emerged later, **the early-stage stability suggested the need to refine the modelling of spending dynamics during that period.** This issue was subsequently resolved by the OpenDP team through an adjustment to the model

**3.2.11 Correlation Analysis Results**

- Positive Findings:
  The cross-correlation analysis showed promising results, with most lags clustering around 0–3 weeks, indicating a strong and timely relationship between epidemiological data and transactional patterns:

```
> df_ccf_abs_max %>% arrange(admin_name1) %>% print(n = 50)
# A tibble: 40 × 4
   merch_category                        admin_name1 lag_max ccf_max
   <chr>                                 <chr>         <dbl>   <dbl>
 1 Airlines                              Bogota            1   0.745
 2 Bars/Discotheques                     Bogota            1   0.925
 3 Computer Network/Information Services Bogota            0  -0.431
 4 Drug Stores/Pharmacies                Bogota            1  -0.562
 5 General Retail Stores                 Bogota            1   0.739
 6 Grocery Stores/Supermarkets           Bogota            2   0.371
 7 Hospitals                             Bogota            0  -0.476
 8 Hotels/Motels                         Bogota            2   0.757
 9 Restaurants                           Bogota            1   0.885
10 Utilities: Electric, Gas, Water       Bogota           34   0.149
11 Airlines                              Brasilia          2   0.752
12 Bars/Discotheques                     Brasilia          1   0.867
13 Computer Network/Information Services Brasilia          0  -0.205
14 Drug Stores/Pharmacies                Brasilia          2  -0.540
15 General Retail Stores                 Brasilia          1   0.699
16 Grocery Stores/Supermarkets           Brasilia          3   0.449
17 Hospitals                             Brasilia          1  -0.410
18 Hotels/Motels                         Brasilia          1   0.720
19 Restaurants                           Brasilia          2   0.851
20 Utilities: Electric, Gas, Water       Brasilia         13  -0.206
21 Airlines                              Medellin          1   0.585
22 Bars/Discotheques                     Medellin          1   0.849
23 Computer Network/Information Services Medellin         55   0.259
24 Drug Stores/Pharmacies                Medellin          1  -0.505
25 General Retail Stores                 Medellin          2   0.653
26 Grocery Stores/Supermarkets           Medellin          0   0.385
27 Hospitals                             Medellin          0  -0.481
28 Hotels/Motels                         Medellin          0   0.699
29 Restaurants                           Medellin          1   0.813
30 Utilities: Electric, Gas, Water       Medellin         21  -0.216
31 Airlines                              Santiago         86   0.323
32 Bars/Discotheques                     Santiago         86   0.445
33 Computer Network/Information Services Santiago         19  -0.169
34 Drug Stores/Pharmacies                Santiago         48  -0.209
35 General Retail Stores                 Santiago         86   0.371
36 Grocery Stores/Supermarkets           Santiago         13  -0.250
37 Hospitals                             Santiago         27   0.263
38 Hotels/Motels                         Santiago         86   0.321
39 Restaurants                           Santiago         86   0.441
40 Utilities: Electric, Gas, Water       Santiago         53  -0.267
```

Figure 3.6: Comparison of Cross-Correlation Lags between Financial Transactions and Epidemiological Metrics (Cases vs. Deaths)

- Anomalous Observations:

    - Utilities (Electric, Gas, Water): These categories show unusually high lags (e.g., 34 weeks in Bogota), likely due to the assigned multipliers being 0, indicating weak or no correlation.

    - Hospitals: Negative correlations are observed in 3 of the 4 cities (e.g., Bogota, Brasilia, and Medellin), despite positive multiplier

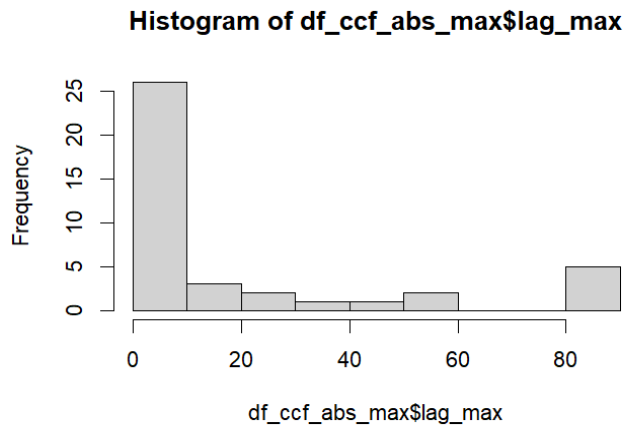

Figure 3.7: Frequency vs Max Lag

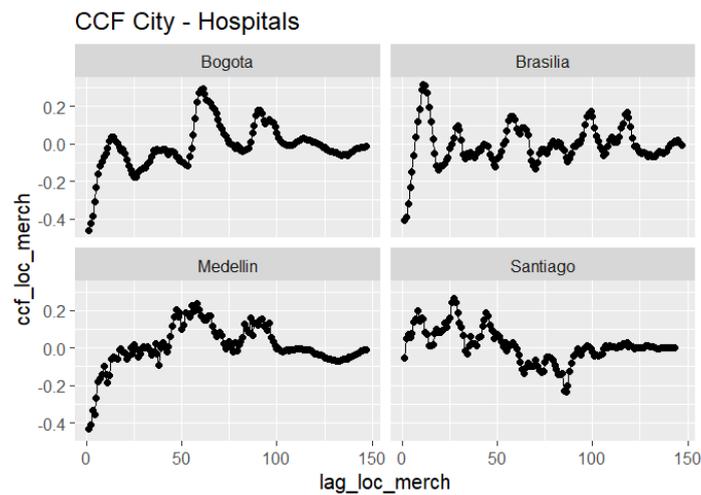

Figure 3.8: CCF vs Lag

We suggest correcting this by switching from cases to deaths as the epidemiological metric.

Recommendations for dataset improvement:

To improve the dataset and its alignment with real-world dynamics, the following steps were recommended:

1. Replace new cases with new deaths as the epidemiological metric to correlate the financial mock dataset with pandemic trends.

2. Adjust transaction sampling proportions based on city populations.

3. Revisit and refine the spending multipliers for each merchant category to reflect dynamic changes during the pandemic.

4. Investigate and resolve anomalies in specific merchant categories, such as Utilities and Hospitals.

These adjustments helped enhance the synthetic dataset's realism and ensure it served as a reliable basis for analysing the interplay between transactional and epidemiological data.

### 3.3 How to apply the use cases to develop applications or solutions from the interpolated & privacy-preserved datasets

Through the challenge, we developed a suite of tools leveraging Privacy-Enhancing Technologies (PETs) to analyze sensitive data to address each of the five use cases outlined *without* compromising individual privacy.

## 4. PETs solutions

Let us now dive deeper into the strategies, analytical approaches, and evaluations of the tools developed.

### 4.1. Tools

#### 4.1.1  Hotspot Detection

- **Purpose:** Enable targeted resource allocation (testing, vaccines, healthcare personnel) to high-risk areas during outbreaks.

- **Methodology:**
  Analyzes financial transactions in a private way, this will serve as a privacy-preserving proxy for physical interactions.

  This algorithm processes transaction data to generate a differentially private visualization of transaction hotspots within a selected city. It begins by enriching the dataset with a city column, mapping each transaction's postal code to its respective city. The data is then filtered to retain only "OFFLINE" transactions, ensuring the analysis focuses on in-person activity.
  Further filtering is applied to isolate transactions occurring within the selected city's postal codes, followed by a temporal constraint to include only transactions within a specified time frame. Next, transactions are aggregated at the postal code level, where a differentially private mechanism applies Gaussian noise to the counts, mitigating the risk of individual re-identification while preserving statistical utility.
  The results are displayed as a colour gradient to highlight transaction hotspots, enabling spatial trend analysis while ensuring privacy compliance.

- **Result:** The tool enhances the speed and efficiency of resource deployment, ultimately leading to a significant reduction in transmission rates and improved public health outcomes. By utilizing anonymized offline transaction data, it can accurately pinpoint areas with high physical presence. This capability enables organizations to allocate resources more effectively, ensuring that interventions are targeted where they are needed most. As a result, the tool plays a crucial role in optimizing response efforts, minimizing the spread of disease, and enhancing overall community well-being.

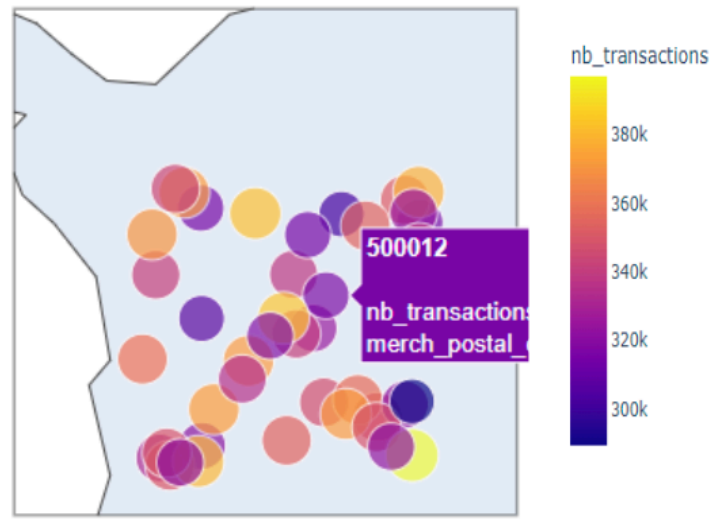

Figure 4.1: Hotspot detection using the DP-epidemiology package. Transaction Location in Medellin from 2020-01-01 to 2020-12-31 with epsilon=1

- **Potential Use Cases:**
    Targeted resource allocation during outbreaks:
    1.1. Identify areas with high presence (e.g., busy markets, public transport hubs, densely populated neighborhoods) to strategically place mobile testing units,
    1.2. Prioritize vaccine delivery to areas identified as high-presence zones,
    1.3. Anticipate surges in demand for healthcare services by identifying high-presence areas
    1.4. Analyze high-presence areas and times to strategically deploy police patrols, emergency medical services

### 4.1.2 Mobility Analysis

- **Purpose:** Inform evidence-based movement policies, optimize resource allocation, and anticipate healthcare demand.

- **Methodology:** Correlates commuting and transportation transactions with mobility trends, validated against existing mobility data.
    This algorithm processes synthetic transaction data to generate a differentially private analysis of mobility trends across different business categories within a selected city. It begins by adding a city column, mapping each synthetic transaction's postal code to its respective city, followed by the addition of a super category column that classifies merchants into retail and recreation, grocery and pharmacy, and transit stations. A time-based filter is applied to restrict transactions to a given time frame. Next, transaction counts are aggregated by postal code for each time step, with Gaussian noise applied to ensure differential privacy.
    The privacy framework incorporates sensitivity and epsilon analysis to balance privacy and data utility. The sensitivity per merchant is set to 3, and for multiple timesteps, it scales as 3 × number of time steps. The epsilon budget is allocated per time step, with the scale parameter calculated as (3 × number of time steps × upper bound) / epsilon. The processed data is validated by comparing the mobility trends derived from synthetic transaction data with external public datasets,

such as the Google COVID-19 Mobility Report for Bogotá, ensuring consistency and reliability of mobility insights while maintaining privacy guarantees.

- **Result:** The tool analyzes synthetic commuting and transportation transactions to identify mobility trends, validating its findings against existing mobility data. By tracking population movements, it enables the prediction of potential case surges, allowing for timely interventions. This data-driven approach supports the development of proactive public health strategies that minimize disruption while maximizing effectiveness. As a result, the tool enhances public health policy optimization, ensuring a more efficient and targeted response to emerging health concerns.

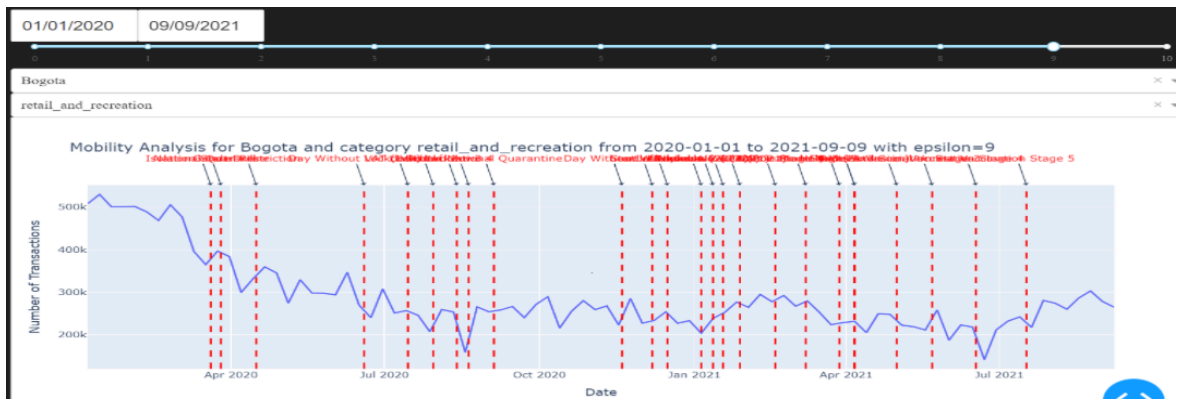

Figure 4.2: Mobility Analysis for Bogota and category=retail_and_recreation

- **Potential Use Cases:**

  1. Targeted Public Health Interventions: Identifying specific neighborhoods or regions with increased mobility despite public health guidelines can help authorities implement targeted interventions like localized testing centers, vaccine drives, or stricter enforcement of mask mandates in those areas.
  2. Evaluating the Effectiveness of Movement Policies: By analyzing mobility data before and after the implementation of policies like lockdowns or travel restrictions, policymakers can objectively assess the impact of these policies on population movement and adjust them accordingly for better effectiveness and reduced societal disruption.

**4.1.3 Pandemic Adherence Analysis-** Gauging Behavioral Shifts & Economic Impact

- **Purpose:** Monitor the effectiveness of public health messaging, understand economic impacts, and inform adaptive policy adjustments.

- **Methodology:** Uses category-specific synthetic transactional data to monitor behavioral shifts and economic trends over time.

  This algorithm processes synthetic transaction data to analyse consumer spending behaviour across essential and luxurious goods while ensuring differential privacy. It begins by adding a city column, mapping each transaction's postal code to its respective city, followed by filtering the dataset to retain only transactions within the selected city. A super category column is then introduced to classify transactions into essential goods (e.g., utilities, drug stores, grocery stores, hospitals, and general retail) and luxurious goods (e.g., hotels, bars, and restaurants).

Next, transactions are aggregated by postal code, with Gaussian noise added to ensure differential privacy, preventing individual transaction identification while preserving overall trends. Both online and offline transactions are considered to provide a comprehensive view.

The synthetic transaction data is visualized to reflect pandemic-stage economic activity, highlighting shifts in essential vs. luxurious spending over time while maintaining privacy safeguards.

- **Result:** The tool provides real-time insights into public behavior and economic resilience, allowing for data-driven policy adjustments. By analyzing spending trends, it helps **assess economic recovery** and monitor public adherence to health policies. It could enable policymakers to make informed decisions, ensuring effective responses to changing public health conditions at large scale.

- **Potential Use Cases**
    1. Policy adherence: Example: After implementing a mandatory mask order, track spending in categories like restaurants, retail, and entertainment to see if it decreases (indicating adherence by reducing social activity) or stays constant/increases (suggesting lower adherence or economic resilience despite the policy
    2. Localized communication: If data shows spending in entertainment venues in a particular neighborhood is consistently high despite public health warnings, targeted messaging campaigns or increased enforcement could be deployed in that area.

### 4.1.4 Contact Matrix Estimation

- **Purpose:** Inform risk assessments for public spaces (schools, workplaces), and guide targeted interventions for vulnerable populations.
- **Methodology:** Infers age-group dynamics through merchandise consumption patterns and derives contact frequencies.

The tool computes **contact patterns** across a country using synthetic transaction data and machine learning-based estimation of age-group-wise consumption behaviour. It first calculates private transaction counts at the city level, then derives private counts per age group using an estimated merchandise consumption distribution (D).

A machine learning approach iteratively refines D by minimizing the difference between the estimated and ground truth contact matrices. This requires country-specific training, assuming the availability of a ground truth matrix aligned with the transaction data. It computes age-group contact counts for each city and averages them to form a national contact matrix. To ensure symmetry and account for age-dependent mixing patterns, the matrix is adjusted using a mixing factor vector and averaged with its transpose.

- **Result:** The tool estimates contact matrix data for epidemiological analysis by leveraging synthetic transactional data to assess interaction rates between age groups and understand transmission dynamics. This data-driven approach enables improved epidemiological analysis.

- **Potential Use Cases:**
    1. Informing Epidemiological Models: Contact matrices are crucial inputs for epidemiological models (like SIR, SEIR models) that predict the spread of infectious diseases. This tool can provide *real-world, data-driven* contact rates

between age groups, making these models more accurate and reliable

## 4.2. Frameworks

### 4.2.1 Outbreak Detection Framework

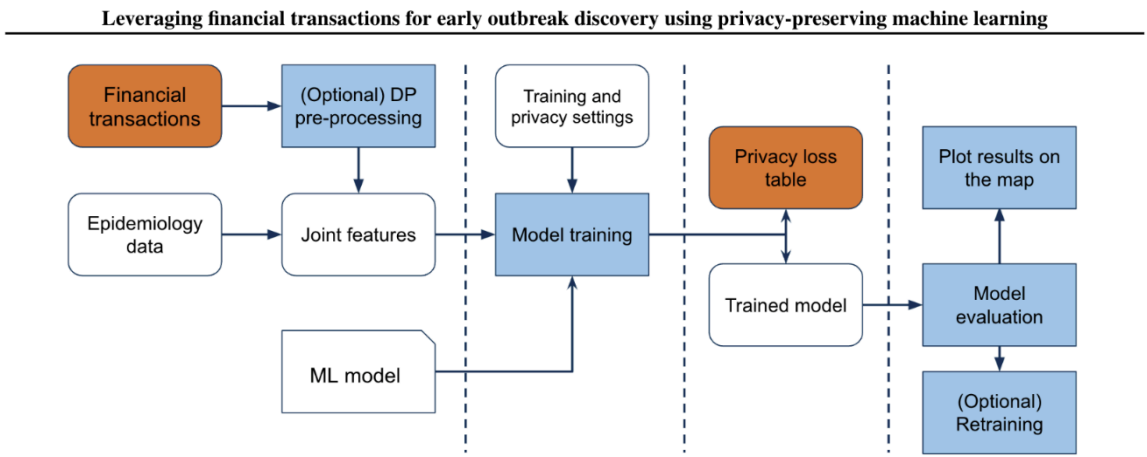

Figure 4.3: Overview of proposed framework

- **Purpose:** Provide policymakers and early responders with critical early warning, enabling proactive resource allocation and containment measures.

- **Methodology:** Leverages ML on combined datasets (transactional and epidemiological knowledge) to identify anomalous patterns signaling early outbreaks under privacy-constrained settings by leveraging differential privacy (in the form of DP-SGD or DP input data pre-processing). [7]

- **Result:**
  This framework for the early detection of outbreaks leverages real-time transaction data using differential privacy. Designed to work with various epidemiological models and differential privacy approaches, this framework enables proactive intervention and detection.
  It is widely adoptable which ensures broad applicability across different public health scenarios. While the framework was evaluated on the dataset described above, there is no specific requirement to the input data with an exception of an identifier similar to merchant_id used to track the privacy loss with respect to individuals.

### 4.2.2 Beyond the tools - DPEpiNN and EpiDP Frameworks

- **DPEpiNN (Deep Privacy-Preserving Epidemic Neural Network):** A novel joint neural network and epidemic model integrating heterogeneous datasets with differential privacy. Enables robust forecasting and epidemic characteristic estimation.

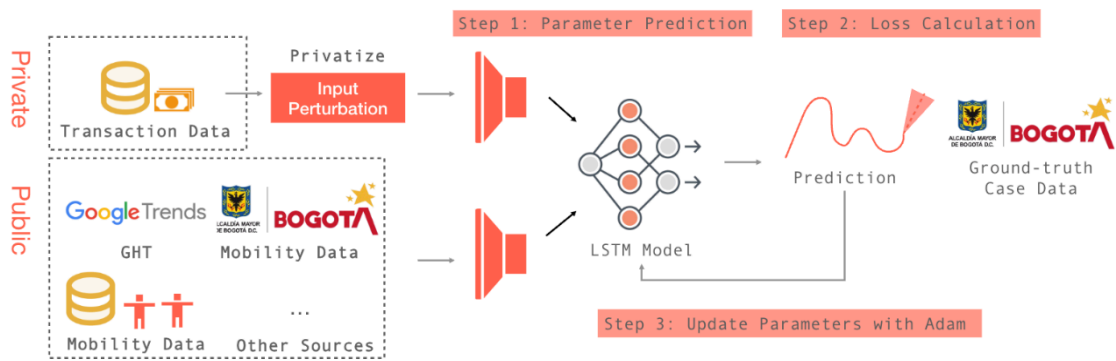

Figure 4.4: Forecasting using the LSTM technique: we use a multi-encoder structure, which allows different types of datasets (whose shapes are not always consistent) to be incorporated. Transaction data is incorporated through a separate encoder. We have considered different strategies for including the Google mobility data—as part of a fixed combination with other datasets, or through a separate encoder, whose weights are learned. The predictions are directly generated by the LSTM's outputs

- **EpiDP (Epidemiological Differential Privacy Package):** An open-access, user-friendly environment for incorporating covariate time series (like mobility and transaction data) to improve real-time Rt estimates and assess epidemiological value.

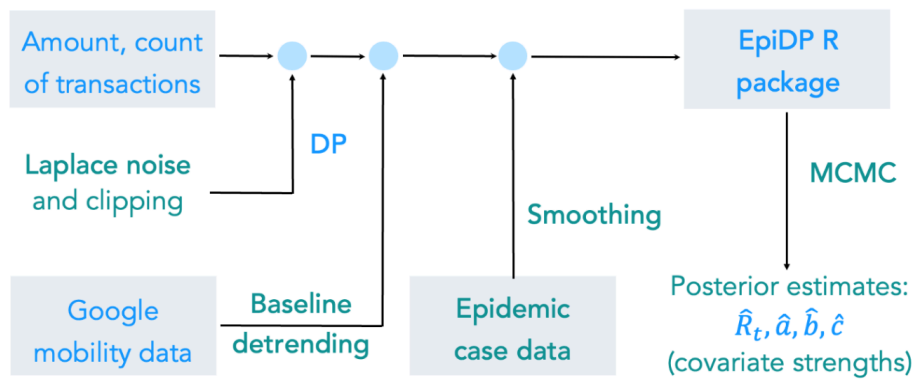

Figure 4.5 EpiDP functionality. Information from differentially private (DP) transaction data and anonymised mobility data can both be appended to case data to generate improved and more robust Rt estimates. The relative importance of these additional data sources is also provided via the estimated regression coefficient strengths (a, b, c)

- **Key Benefits:**
    1.1. **Data Integration:** Handles diverse data types (transaction, mobility, etc.)
    1.2. **Privacy by Design:** Differential privacy ensures robust privacy guarantees.
    1.3. **Flexibility & Extensibility:** Adaptable to new data sources and epidemic challenges.
    1.4. **Open Access:** Promotes broader adoption and further innovation in the field.

# 5. Evaluation of PETs Tools

Each tool/framework was evaluated for its innovative potential and practical application and summarized as follows:

5.1 Multiscale PET data metrics for improved early detection and response to epidemics: the need for better integration of financial data with epidemiological metrics and suggested improvements in Bayesian modelling to enhance the estimation of Rt.

5.2 Privacy-enhanced models for transactional data: the need to improve hotspot detection and mobility analysis, ensuring the methods were validated against external datasets for reliability in real-time scenarios.

5.3 Privacy-preserving anomaly detection for early outbreak discovery: the need to refine anomaly detection methods by incorporating standard epidemiological practices and clarifying the criteria used to define anomalies.

5.4 Joint deep learning and epidemic transmission model for public health analyses with differential privacy: the need to restructure the tool for improved usability and scalability, enhancing its visualization capabilities, and expanding the documentation for better user guidance.

**Detailed Evaluation**

5.1 Multiscale PET data metrics for improved early detection and response to epidemics

*Imperial College London (ICL) and University of Oxford*

The tool is an R package that interfaces with Stan to implement a Bayesian model based on the renewal equation for estimating the effective reproduction number (Rt) using external covariates besides the typical incidence data. For example, assuming a sinusoidal function with Gaussian noise for the Rt (this kind of emulates successive epidemics), and using the simulated incidence as an input for the Bayesian model, they manage to recover the Rt from which they simulated with good accuracy. Another example showcases how to use mobility data as a covariate in the Bayesian model.

Despite the effort to use financial transactions as covariates to improve Rt estimations, the results show no significant improvement concerning the case where mobility trends are used as covariates instead, as exemplified by Figure 5.1:

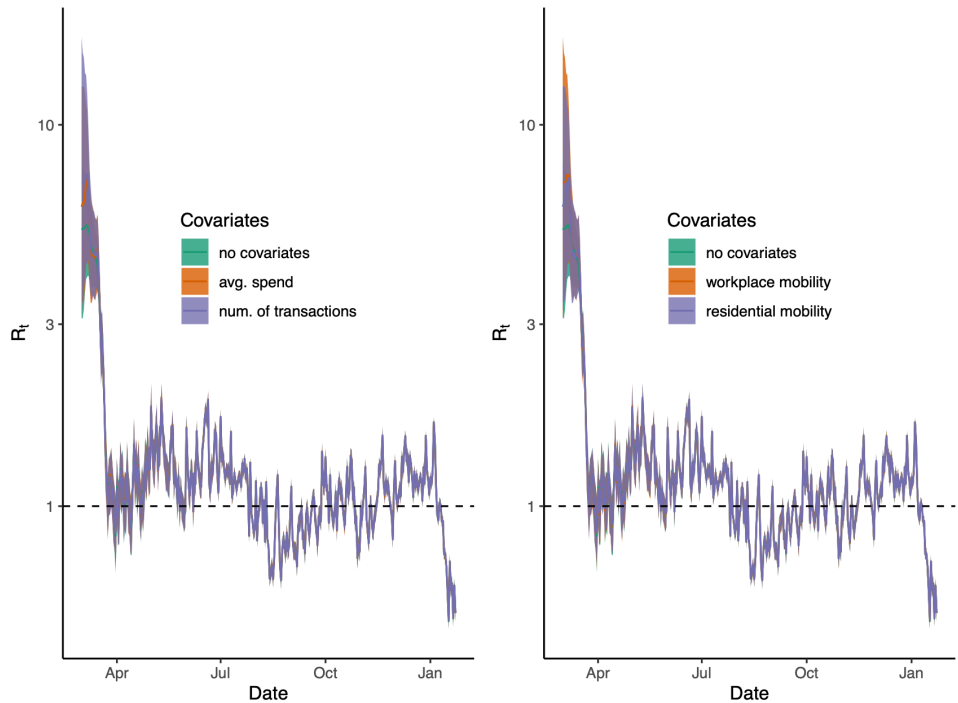

Figure 5.1: Rt estimation comparison using financial transactions (left) and mobility trends (right) as covariates in the Bayesian model. The underlying code for R_t estimation was adapted from methodology in [8].

## 5.2 Privacy-enhanced models for transactional data

### Indian Institute of Technology (IIT) Kanpur

This team designed a Python package intended for implementing differential private release of financial data to be used in an epidemiological context. The package can be used to construct the following analysis:

- **Hotspot Detection:** This analysis tracks pandemic hotspots by monitoring differential private release of financial transactions in a city and identifying areas with high transaction activity.

- **Mobility Detection:** This analysis tracks mobility by monitoring differential private time series release of financial transactions in the retail and recreation, grocery and pharmacy and transit stations super categories, which should match with Google mobility data for easy validation.

- **Pandemic Adherence Detection:** They analyze transaction behavior to identify pandemic stages by comparing transactions in essential vs luxurious goods categories.

- **Contact Pattern Matrix Estimation:** Estimates the contact matrix by analyzing transactional data for different age groups across various merchandise categories.

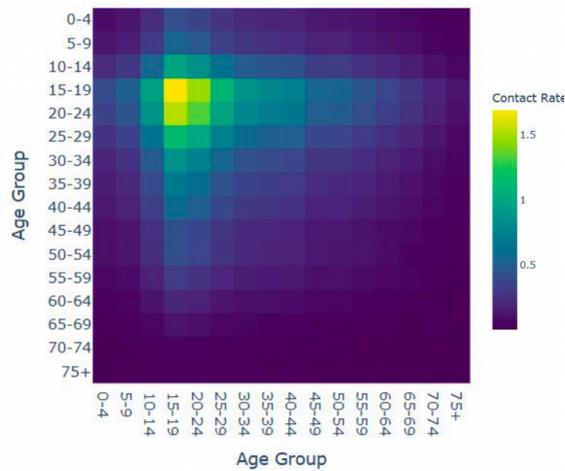

Figure 5.2: Estimated contact matrix using the DP-epidemiology package.

For a full analysis of this approach, refer to the corresponding package documentation [9]

5.3 Privacy-preserving anomaly detection for early outbreak discovery [10]

*Technical University of Munich (TUM)*

This proposed solution supports two types of users: the informed and the uninformed. The former has some understanding of what an early outbreak may look like: this can be represented in a sharp rise in the **R** number, specific threshold of new cases over a given time period etc. To support this type of user the proposed solution relies on supervised ML training, where the user themselves can specify what constitutes an early outbreak using binary labelling (i.e. 1 for early outbreak and 0 otherwise) [11] [12]. This allows them to leverage their additional knowledge in order to rely on the most well-tested and well-supported part of the framework that concentrates on the fully labelled data.

The latter user type, however, does not possess the knowledge to label the individual data points and simply wants to learn as much as possible from a given dataset. For this user type the solution employs unsupervised anomaly detection. Specifically an autoencoder-based approach in order to leverage the statistical properties of the dataset and identify which specific data points fall on the tails of the input data distribution (or belong to a different distribution altogether). This user can benefit from identifying the anomalies in the data without having any prior knowledge about the data and only having some relatively broad understanding of statistics.

The approach demonstrates strong performance under both user types in the tasks, with a stronger performance in the informed use case (which was expected). The comparison between two approaches can be found in Figure 5.4.

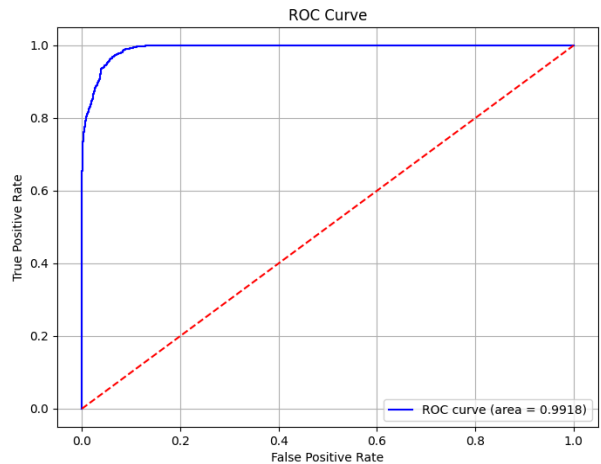

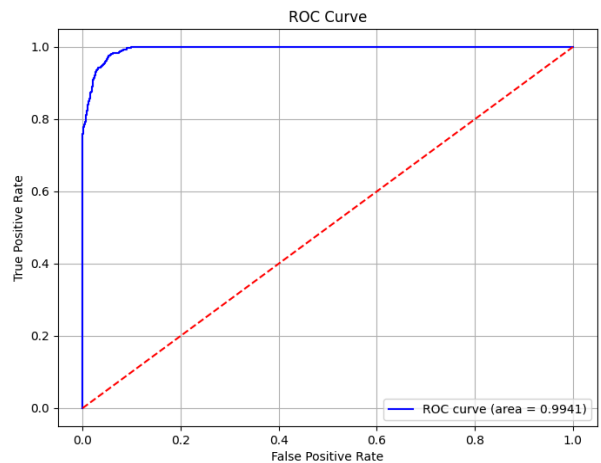

Figure 5.3: Using DP-SGD in the supervised classification setting (i.e., for an informed user) with ε = 1.0 (top) and ε = 4.0 (bottom). Both models achieve excellent performance, even under strict privacy constraints.

At the current stage of development the framework is capable of training any model–dataset combination for anomaly detection in both supervised (binary) and unsupervised (autoencoder-based) settings. The framework is flexible and allows the user to work with a variety of privacy regimes, including individual differential privacy in its basic form with privacy filtering, as well as continuous retraining if the user maps the previously trained model to the privacy loss table. In a supervised setting it is possible to achieve close to 100% accuracy at ε = 8.0 (which should be interpreted cautiously given the dataset size), demonstrating that it is possible to train an effective model using this framework.[13][14]

However, the main challenge, which is beyond the scope of this project, is finding suitable data. While the framework supports any compatible model–dataset combination, this still relies on the user identifying these combinations in advance. An uninformed user may still find the framework valuable, but the quality of results is unlikely to match those obtained by an informed user. Such a model would also likely face broader generalisation issues, as retraining on a different dataset without a clear objective can lead to highly unpredictable performance (since an autoencoder-based model may not fit data originating from a substantially different distribution).

Therefore, the most effective use of this framework is through strong collaboration between epidemiologists and data owners, where epidemiologists request data suited to their learning objectives and data owners prepare it in the correct format in advance.

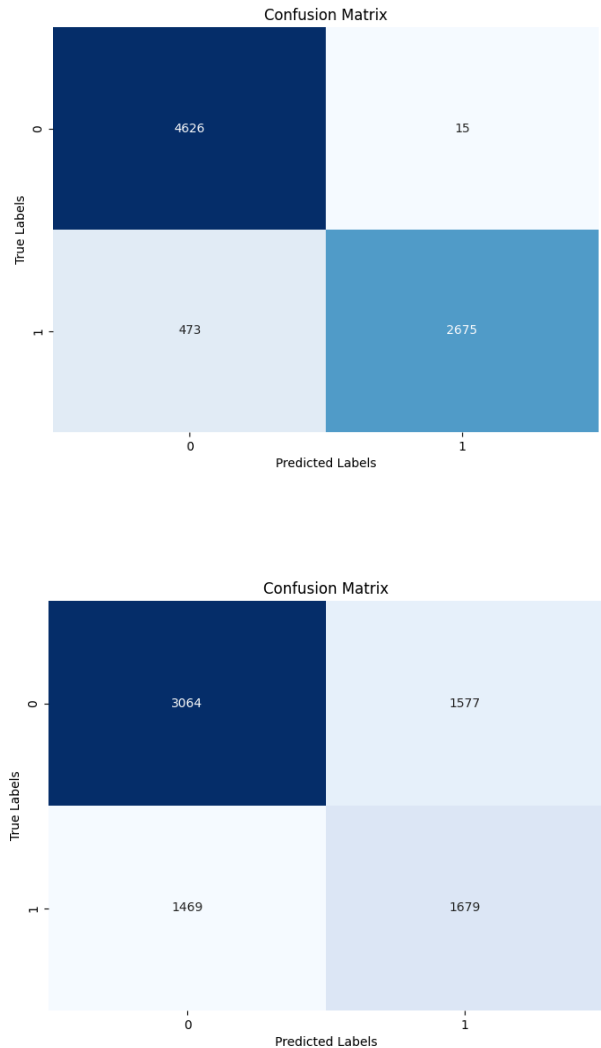

Figure 5.4: Comparison of the *informed* (top) and the *uninformed* (bottom) user's utility in the presence of pre-existing labels (i.e. the training data is different, but the evaluation data is the same and contains labelled anomalies for a fairer comparison).

## 5.4 Joint deep learning and epidemic transmission model for public health analyses with differential privacy

*University of Virginia (UVA), Georgia Tech, and the University of Arizona*

The tool trains a joint neural network that estimates the parameters of a set of possible compartmental models. Once this has been done, they can do forecasting using standard machine learning methods.

The proposed framework is said to support a broad class of epidemic analyses using diverse types of datasets, with differential privacy. These analyses are of two different types. The first consists of forecasting and nowcasting, for which they use a Long Short-Term Memory (LSTM) method. The second involves the estimation of epidemic characteristics and evaluation of interventions, which require learning and using an epidemic model. [15]

Using this framework, they performed a counterfactual analysis of a social distancing type intervention which is said to lead to a reduction in transmission rate by a given amount, starting at a fixed time.

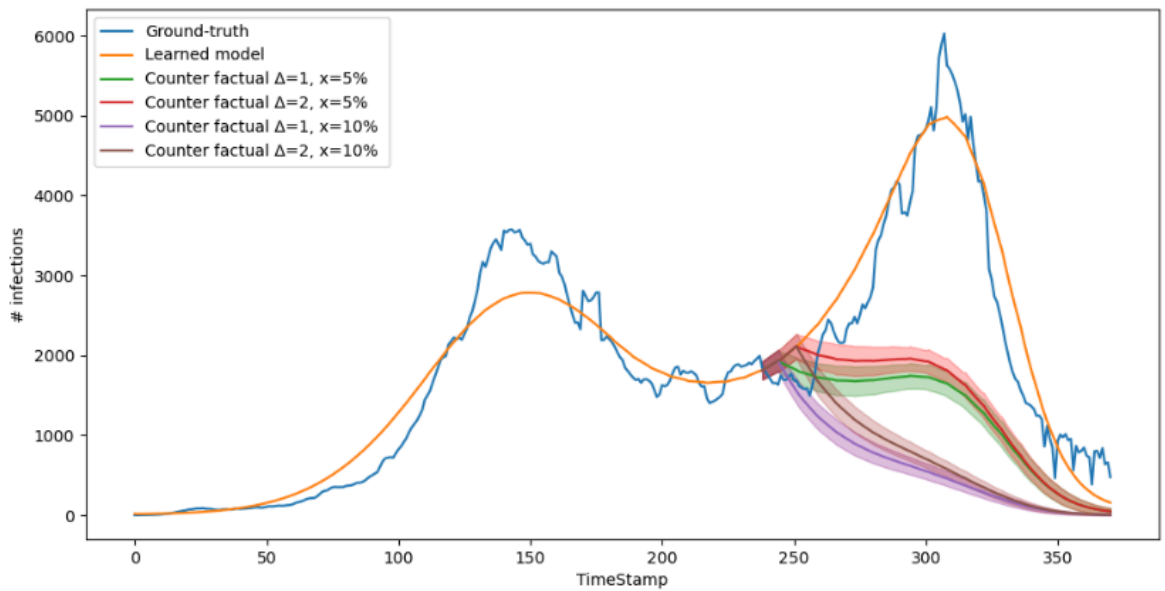

Figure 5.5: UVA's counterfactual analysis.

For a full analysis of this approach, refer to the corresponding paper [16]

# 6. Conclusion

In conclusion, the PETs for Public Health Challenge has established a robust framework for integrating privacy-enhancing technologies into epidemiology. The tools and methodologies developed through this initiative provide a valuable roadmap for future advancements, emphasising the importance of balancing data utility with privacy to achieve impactful public health outcomes. Indeed, the findings and evaluations presented throughout this document highlight several key outcomes and implications:

- Multiscale PET data metrics for improved early detection and response to epidemics: We have developed a Bayesian framework using synthetic financial data and epidemiological data to refine estimates of the effective reproduction number (Rt). While mobility data was effectively used as a covariate, integrating synthetic financial transactions highlighted the potential for economic activity metrics to inform disease transmission patterns. This approach underscores the value of incorporating diverse data sources for more accurate epidemiological modelling.

- Privacy-enhanced models for transactional data – We have introduced a Python package for hotspot detection, mobility analysis, and contact pattern estimation using financial transaction data. The tool's focus on privacy-preserving analysis provides a foundation for real-time decision-making and public health interventions, particularly through its innovative use of synthetic transaction data as a proxy for social mobility and interaction patterns.

- Privacy-preserving anomaly detection for early outbreak discovery- We have proposed a privacy-preserving anomaly detection framework that combines synthetic financial data and epidemiological data to identify early outbreak signals. This approach is notable for its potential to serve as an early warning system, enhancing the timeliness and effectiveness of public health responses through predictive modelling.

- Joint deep learning and epidemic transmission model for public health analyses with differential privacy – We have developed a neural network-based tool for forecasting and counterfactual analysis. By integrating compartmental models with differential privacy, the framework supports a broad range of analyses, including nowcasting and evaluation of public health interventions. The use of advanced machine learning techniques highlights the growing role of artificial intelligence in epidemiological research.

It is important to highlight that the innovative approaches developed through this challenge provide a roadmap for future advancements in privacy-preserving public health analytics. Continued efforts should focus on refining these methodologies, enhancing data integration capabilities, and fostering interdisciplinary collaboration to address emerging public health challenges and extend these tools to new use cases and domains (as suggested above).

Finally, our project has shown that the integration of PETs in pandemic response represents a paradigm shift in balancing data utility with privacy. By leveraging differential privacy and innovative analytical models, these technologies provide secure and actionable insights for public health and beyond. Scaling these solutions across various domains could revolutionize data-driven decision-making in an increasingly interconnected world.

Further work is needed to build and deploy the above approaches at a larger scale to continue to test and develop these ideas.

Epidemic Analysis. arXiv Preprint. arXiv:2506.22342.